\documentclass[aps,twocolumn,pra,superscriptaddress,showpacs,tightenlines]{revtex4}
\usepackage{amssymb}
\usepackage{amsmath}
\usepackage{graphicx}
\usepackage{epsfig}
\usepackage{subfigure}
\usepackage{amsfonts}
\usepackage{CJK}

\begin{document}

\begin{CJK*}{GBK}{song}
\title{Bloch oscillations of quasispin polaritons in a magneto-optically controlled atomic ensemble}

\author{Chang Jiang}
\affiliation{Key Laboratory of Low-Dimensional Quantum Structures
and Quantum Control of Ministry of Education, and Department of
Physics, Hunan Normal University, Changsha 410081, China}
\author{Jing Lu}
\thanks{Corresponding author}
\email{lujing@hunnu.edu.cn}
\affiliation{Key Laboratory of Low-Dimensional Quantum Structures
and Quantum Control of Ministry of Education, and Department of
Physics, Hunan Normal University, Changsha 410081, China}
\author{Lan Zhou}
\affiliation{Key Laboratory of Low-Dimensional Quantum Structures
and Quantum Control of Ministry of Education, and Department of
Physics, Hunan Normal University, Changsha 410081, China}

\begin{abstract}
We consider the propagation of a quantized polarized light in a
magneto-optically manipulated atomic ensemble with a tripod
configuration. Polariton formalism is applied when the medium is
subjected to a washboard magnetic field under electromagnetically
induced transparency. The dark-state polariton with multiple
components is achieved. We analyze quantum dynamics of the
dark-state polariton by some experiment data from rubidium D1-line.
It is found that one component propagates freely, however the
wavepacket trajectory of the other component performs Bloch
oscillations.
\end{abstract}
\maketitle \end{CJK*}\narrowtext 
\narrowtext

\section{\label{Sec:1}Introduction}

Quasiparticles are regarded as collective excitations of many
elementary particles, as well as the mixtures of different
elementary excitations, which are the basic constructions of matters
together with elementary particle. Quasiparticles are crucial for
understanding many phenomena in condensed matter
physics~\cite{Firstenberg,Firstenberg1}. One of the
fundamental phenomena in condensed matter physics is the dynamics of
particles in a periodic potential under the influence of a static
force. As it is well-known that a quantum particle in a periodic potential
possesses energy eigenvalues forming Bloch bands and delocalized eigenstates
known as Bloch functions, the particle undergoes uniform motion. After
a static force is applied, the eigenstates become localized, and the
energy spectrum becomes discrete with the formation of Wannier-Stark
ladders~\cite{Wannier60,Wacker02,Korsch02}. Contrary to common sense,
this particle oscillates instead of infinitely accelerating by the force,
i.e., the famous Bloch oscillation~\cite{Bloch28,Zener34}. In physics,
such an oscillation is generally explained as a Bragg reflection of
the accelerated particle, which causes a wave packet to oscillate
rather than translate through the lattice.

With the advances of atomic physics and quantum optics over the last
decades, considerable attention has been paid on quasiparticles of
light-matter interaction since they are suggested as new systems to
simulate a variety of many-body phenomena~\cite{Plenio08,{Fazio10}}
with unprecedented precision and control. A prominent phenomenon in
light-matter interaction is electromagnetically induced transparency
(EIT), where the transmission of a probe beam through an optical
dense medium is manipulated by means of a control
beam~\cite{HarrisTD,RMP77(05)633}. Stimulated by the construction of
both quantum memories and quantum carriers free of quantum
decoherence, dark state polaritons emerge in storing and
transferring quantum state of light to collective atomic excitations
of matter in EIT. A dark state polariton (DSP) is a quasiparticle,
which is a bosoniclike collective excitation of a photon and an
atomic spin wave~\cite{SunPRL03,RMP75(03)457}. The particle nature
of dark state polaritons, possessing an effective magnetic moment,
has been demonstrated by the enhanced deflection of the laser beam
after light propagates through a $\Lambda$-type atomic vapor with a
magnetic-field gradient applied~\cite{PRL78(97)003451,Karpa06}, and
an effective Schr\"{o}dinger equation is derived to exhibit the
wave-particle duality of the dark state polaritons~\cite{ZhouA08}.
Bloch oscillations of single DSP are proposed~\cite{ZhangA10}, and a
method is described to create an effective gauge potential for a
DSP~\cite{104(10)033903}.

Another remarkable property of DSPs in EIT is the presence of
multiple ``spin'' components, which open up the possibility to study
a variety of many-particle effects in effective magnetic fields. A
phenomenon of birefringence in EIT has been predicted as a
generalized Stern-Gerlach effect of quantized polarized
light~\cite{yuguoA08,ZhangA09}. Collapses and Revivals of DSPs are
also experimentally demonstrated~\cite{KuzmichA06,KuzmichL06}. Here,
we study spatial motion of the DSP with two components in two
inhomogeneous magnetic fields consisting of a periodical magnetic
field and a magnetic field gradient. The magnitudes of the magnetic
fields vary in a direction transverse to the incident direction of
the probe beam. This investigation is inspired by the following
technical advantages of DSPs: refractive index modulation
straightforwardly creates a scalar potential; a direct measurement
is simple for photons; and the waveguide and resonator techniques
can be used to confine the spatial motion of polaritons to lower
dimensions. In this paper, the DSP with two components is generated
in an atomic ensemble with a tripod configuration, which controlled
by a specially designed magneto-optical manipulation based on the
EIT mechanism. After obtaining the equation of motion for this
quasispin DSP by the perturbation theory, we employ the single-band
and tight-binding model to give an analytic treatment of the
dynamics of DSP. It is found that one DSP component acts as a free
particle, the other component experiences a harmonic motion with
angular frequency proportional to the steady force and its momentum
growing linearly in time (i.e. Bloch oscillations). The experimental
feasibility of the proposed scheme is also discussed using rubidium
87 D1-line.

This paper is organized as follows. Section~\ref{Sec:2} describes the
theoretical model for four-level atoms with a tripod configuration in
external fields. In Sec.~\ref{Sec:3}, the quasispin DSP has been introduced
as dressed fields to describe the spatial motion of collective
excitation. Afterward, we present the evolution equation for this quasispin
DSP. In Sec.~\ref{Sec:4}, we present the results of detailed calculations
for the propagation of the DSP in a washboard magnetic field within
single-band and tight-binding approximation. We conclude our paper in the
final section.

\section{\label{Sec:2} Theoretical model for tripod atomic ensemble in
external fields.}

\begin{figure}[tbp]
\includegraphics[bb=52 435 546 665, width=8 cm]{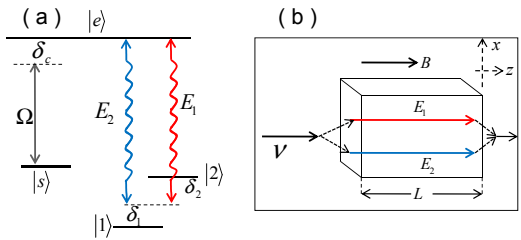}
\caption{(Color online) Tripod scheme of matter-light interaction
involving a linear-polarized probe field and one control field in
the presence of a magnetic field parallel to the field propagation
direction.} \label{fig2:1}
\end{figure}

We deal with an ensemble of 2N identical and noninteracting atoms. Each atom
is characterized by three lower states $\{|1\rangle,|2\rangle,|s\rangle\}$,
and an excited state $|e\rangle $, as schematically shown in Fig.~\ref{fig2:1}.
The two ground states $|1\rangle $ and $|2\rangle$ are the degenerate Zeeman sublevels.
Atoms interact with a linear-polarized probe field with frequency $\nu $ and wave
vector $\mathbf{k}$ propagating in the positive $z$ direction, and a classical
control light with frequency $\nu _{c}$ and wave number $\mathbf{k}_{c}$. Here
$k$ and $k_{c}$ are the wave numbers to the central frequencies $\nu $ and
$\nu _{c}$ of the probe and control fields, respectively. The excited state
$\left\vert e\right\rangle $ with $m_{F}^{e}=0$ is coupled to the ground state $%
\left\vert 1\right\rangle $ ($\left\vert 2\right\rangle $) with $%
m_{F}^{1}=-1 $ ($m_{F}^{2}=1$) via a $\sigma ^{+}$ ($\sigma ^{-}$)
component $E_{1}$($E_{2}$)~\cite{Haun409,Lukin86}. The control field
with Rabi frequencies $\Omega $ is assumed to be uniform throughout
the sample, and it drives the transition $|s\rangle \rightarrow
|e\rangle $ with detuning $\delta _{c}$, which create transparent
windows for two components of the probe field. We assume that the
detuning of component $E_{1}$($E_{2}$) from its corresponding atomic
transition is caused by the applied magnetic fields along the $z$
axis with the corresponding amount $\delta _{i}=\mu _{i}B$, the
magnetic moments $\mu _{i}=m_{F}^{i}g_{F}^{i}\mu _{B}$ are
determined by the Bohr magneton $\mu _{B}$, the gyromagnetic factor
$g_{F}^{i}$ and the magnetic quantum number $m_{F}^{i}$ of the
corresponding state $\left\vert i\right\rangle $.

We now introduce components $E_{j}(\mathbf{r},t)$ of the probe field which
vary slowly in space and time~\cite{Lukin1}
\begin{equation}
\tilde{E}_{j}^{+}(\mathbf{r},t)=\sqrt{\frac{\nu }{2\varepsilon _{0}V}}E_{j}(%
\mathbf{r},t)e^{i(kz-\nu t)},(j=1,2)
\end{equation}%
and slowly varying operators
\begin{subequations}
\label{M2}
\begin{eqnarray}
\sigma _{es} &=&\tilde{\sigma}_{es}\exp (ik_{c}z)\text{,} \\
\sigma _{ej} &=&\tilde{\sigma}_{ej}\exp (ikz),(j=1,2)\text{.}
\end{eqnarray}%
Here continuous atomic-flip $\tilde{\sigma}_{\mu \nu }\left( \mathbf{r}%
,t\right) =(1/N_{r})\sum_{r_{j}\in N_{r}}\tilde{\sigma}_{\mu \nu }^{j}\left(
t\right) $ is averaged over a small but macroscopic volume containing many
atoms $N_{r}=(2N/V)dV\gg 1$ around position $\mathbf{r}$, where $2N$ is the
total number of atoms, $V$ is the volume of the medium~\cite{Lukin,Lukin1}%
, and $\tilde{\sigma}_{\mu \nu }^{j}\left( t\right) =\left\vert \mu
\right\rangle _{j}\langle \nu |$ is the flip operator of the jth atom. Above
and hereafter, we take $\hbar =1$. Under the electric-dipole approximation
and the rotating-wave approximation, the interaction Hamiltonian reads
\end{subequations}
\begin{eqnarray}
H_{I} &=&\frac{2N}{V}\int d^{3}r\left[ \left( \delta _{1}\sigma _{11}+\delta
_{2}\sigma _{22}+\delta _{c}\sigma _{ss}\right) \right.  \notag  \label{M3}
\\
&&\left. -\left( \Omega \sigma _{es}+gE_{1}\sigma _{e1}+gE_{2}\sigma
_{e2}+h.c.\right) \right] ,
\end{eqnarray}%
in a frame rotating with respect to the probe and driving fields.
The parameter $g=\langle e|d|1\rangle \sqrt{\nu /(2\varepsilon
_{0}V)}$ characterizes the strength of coupling between the probe
field and the atoms. Due to the symmetry of the states $|1\rangle $
and $|2\rangle $, the transition matrix elements in the above
equation are equal. Although we assume three-photon resonance in
the absence of the magnetic field, that is, $\nu =\omega
_{e1}^{0}=\omega _{e2}^{0}$, $\nu _{c}=\omega _{es}^{0}$, where
$\omega _{e1}^{0}=\omega _{e2}^{0}$ and $\omega _{es}^{0}$ are the
atomic resonance frequencies without the applied magnetic field, we
note that the interaction Hamiltonian $H_{I}$ is the general expression
in the interaction picture for a tripodlike linkage pattern. Here,
$\delta _{i}$ $(i=1,2,c)$ is the detuning between the laser
field and its corresponding atomic transition.

\section{\label{Sec:3}Evolution equation for the DSP with two components.}

When a light pulse enters a medium, photons interact with atoms of the
medium. They are converted into composite quasiparticles of the radiation
and atomic excitations known as polaritons. In EIT system, there are two
types of polaritons ---- the DSP and the bright state polariton. To obtain
the equation that describes the dark-state polariton, first we deal with
the dynamics of this atomic ensemble. The Heisenberg equations read
\begin{subequations}
\begin{eqnarray}
{\dot{\sigma}_{21}} &=&\left[ i\left( \delta _{2}-\delta _{1}\right) -\gamma
_{c}\right] \sigma _{21}+ig^{\ast }E_{1}^{+}{\sigma _{2e}}-ig^{\ast
}E_{2}^{+}\sigma _{e1}, \\
{\dot{\sigma}_{e1}} &=&\left( -i\delta _{1}-\frac{\Gamma }{2}\right) \sigma
_{e1}-ig^{\ast }E_{1}^{+}\left( \sigma _{11}-\sigma _{ee}\right) -ig^{\ast
}E_{2}^{+}\sigma _{21}  \notag \\
&&-i\Omega ^{\ast }\sigma _{s2}, \\
{\dot{\sigma}_{s1}} &=&\left[ i\left( \delta _{c}-\delta _{1}\right) -\gamma
_{c}\right] \sigma _{s1}+ig^{\ast }E_{1}^{+}\sigma _{se}-i\Omega ^{\ast
}\sigma _{e1}, \\
{\dot{\sigma}_{e2}} &=&\left( -i\delta _{2}-\frac{\Gamma }{2}\right) \sigma
_{e2}-ig^{\ast }E_{2}^{+}\left( \sigma _{22}-\sigma _{ee}\right) -ig^{\ast
}E_{1}^{+}\sigma _{12}  \notag \\
&&-i\Omega ^{\ast }\sigma _{s2}, \\
{\dot{\sigma}_{s2}} &=&\left[ i\left( \delta _{c}-\delta _{2}\right) -\gamma
_{c}\right] \sigma _{s2}+ig^{\ast }E_{2}^{+}\sigma _{se}-i\Omega \sigma
_{e2}, \\
{\dot{\sigma}_{se}} &=&\left( i\delta _{c}-\frac{\Gamma }{2}\right) \sigma
_{se}+ig^{\ast }E_{1}\sigma _{s1}-ig^{\ast }E_{2}\sigma _{s2}  \notag \\
&&-i\Omega \left( \sigma _{ee}-\sigma _{ss}\right) ,
\end{eqnarray}
\end{subequations}
where we have phenomenologically introduced the coherence relaxation
rate $\gamma _{c}$ of the ground states and the decay rate $\Gamma$
of the excited state. Here, we consider the case where the intensity
of the quantum field is much weaker than that of the coupling field,
and the number of photons contained in the signal pulse is much less
than the number of atoms in the sample. Therefore, the perturbation
approach can be applied to the Heisenberg equations of the atomic
part of the order in $gE_{j}$, which is introduced in terms of
perturbation expansion
\begin{equation}
\sigma_{\mu\nu}=\sigma^{(0)}_{\mu\nu}+\lambda\sigma^{(1)}_{\mu\nu}+\cdots
\end{equation}
In the above equation, $\mu,\nu\in\{1,2,s,e\}$ and $\lambda$ is a
continuously varying parameter ranging from zero to unity. Here,
$\sigma^{(0)}_{\mu\nu}$ is of the zeroth order in $gE_j$ and
$\sigma^{(1)}_{\mu\nu}$ is of the first order in $gE_j$, and so on.
We retain only terms up to the first order in the signal field
amplitude since the linear optical response theory can sufficiently
reflect the main physical features of the spatial motion of the
input pulse with slow group velocity. With the assumption that all
atoms are initially in level $|j\rangle ,(j=1,2)$ without
polarization, i.e., atom $i$ is in a mix state $\rho _{i}=\Sigma
_{j=1,2}|j\rangle \langle j|/2$, we can neglect the population of states $%
|e\rangle $ and $|s\rangle $, as well as the coherence between these
states. Then the first order atomic transition operator $\sigma
_{ij}^{(1)}$ satisfies the following equation
\begin{eqnarray}  \label{DSP1}
\sigma_{je}^{\left( 1\right) }=-\frac{i}{\Omega ^{\ast }}\left( \partial
_{t}-d_{j}\right) \sigma _{js}^{\left( 1\right) },
\end{eqnarray}
which is related to the atomic linear response to the probe field. In the above equation,
\begin{equation*}
d_{j}=i\left( \delta _{j}-\delta_{c}\right)-\gamma _{c},
\end{equation*}

The dark-state and bright-state polaritons are described by the field
operators as
\begin{subequations}
\begin{eqnarray}
\Psi _{j} &=&E_{j}\cos \theta -2\sqrt{N}\sigma _{js}^{(1)}\sin \theta \text{,} \\
\Phi _{j} &=&E_{j}\sin \theta +2\sqrt{N}\sigma _{js}^{(1)}\cos \theta \text{,}
\end{eqnarray}%
\end{subequations}
respectively, which are superpositions of photonic and spin-wave excitations. They are introduced
in the linear regime with respect to the probe field. Here, $\tan \theta =g\sqrt{N}/\Omega $.
The Heisenberg equation for the slowly varying field operator $E_{j}(\mathbf{r},t)$
results in a paraxial wave equation in classical optics
\begin{eqnarray} \label{DSP2}
\left[ i\partial _{t}+ic\partial _{z}+\frac{c}{2k}(\frac{\partial ^{2}}{%
\partial x^{2}}+\frac{\partial ^{2}}{\partial y^{2}})\right] E_{j}=-2g^{\ast
}N\sigma _{je}^{(1)}.
\end{eqnarray}
Here, $c$ is the velocity of light in vacuum. In terms of dark and bright
polariton field operators, equations~(\ref{DSP1}) and~(\ref{DSP2}) read~\cite%
{ZhouA08,yuguoA08}
\begin{eqnarray}
&&\left[ i\partial _{t}+ic\partial _{z}+\frac{c}{2k}\left( \partial
_{x}^{2}+\partial _{y}^{2}\right) \right] \left( \Psi _{j}\cos \theta +\Phi
_{j}\sin \theta \right)   \notag  \label{DSP3} \\
\text{ \ \ \ } &=&i\frac{g^{\ast }\sqrt{N}}{\Omega ^{\ast }}\left( \partial
_{t}-d_{j}\right) \left( \Phi _{j}\cos \theta -\Psi _{j}\sin \theta \right) .
\end{eqnarray}%
Under the condition of EIT, absorption is negligible, the excitation
of the bright-state polariton field $\Phi _{j}$ vanishes
approximately. Then the dynamics of dark-state polariton field $\Psi
_{j}$ is obtained. To show the basic principle of physics, we assume
a sufficiently strong driving field such that $\left\vert \Omega
\right\vert ^{2}\gg \Gamma \gamma _{c}$. In addition, we have let
$\gamma _{c}=0$. To make an analogy to the schr\"{o}dinger equation,
we rewrite the dynamic equation for the two dark-state polaritons as
\begin{eqnarray} \label{EDSP4}
i\partial _{t}\Psi _{j}=\left[
v_{g}P_{z}+\frac{P_{x}^{2}+P_{y}^{2}}{2m}+\mu _{j}^{eff}B(x)\right]
\Psi _{j},
\end{eqnarray}%
Here, effective magnetic moments
\begin{equation*}
\mu _{j}^{eff}=(\mu _{s}-\mu _{j})\sin ^{2}\theta ,
\end{equation*}%
which can be adjusted by the control field. In  addition, the group
velocity $v_{g}=c\cos ^{2}\theta$ along the $z$ direction can be
controlled by the amplitude of the control field. $m=k/{v_{g}}$ is
the effective mass, and $P_{\alpha}=-i\partial _{\alpha}
(\alpha=x,y,z)$ is the momentum operator on the $\alpha$
direction~\cite{yuguoA08}. The steady atomic response in
the applied magnetic fields induces an effective potential for the DSP. Equation~(%
\ref{EDSP4}) indicates that the two components of a DSP behave independently, i.e.
two components of a DSP does not interact with each other. Now, we assume that the
magnetic fields in z direction consist of two components, one is linear with the
expression $B_{1}x $ and the other is periodical in the transverse
$x$ direction. Then the treatment proceeds along the $x-z$ plane.
The evolution equation for the quasispin DSP is rewritten as
\begin{equation*}
i\partial _{t}\Psi _{j}=\left[ v_{g}P_{z}+\frac{P_{x}^{2}}{2m}%
+F_{j}x+V_{j}(x)\right] \Psi _{j},
\end{equation*}%
which means each component $\Psi _{j}$ is subject to a static force $%
F_{j}=\mu _{j}^{eff}B_{1}$ in its corresponding periodic potential $%
V_{j}(x+d)=V_{j}(x)$. Since the effective Hamiltonian along $z$ direction
commutes with that along the transverse direction which refers to
Wannier-Stark Hamiltonian~\cite{Korsch02}, therefore, the investigation is
confined to the transverse direction. We note that the evolution operator
along $z$-axis is equivalent to the translation operator, and the Hermitian
of the Wannier-Stark Hamiltonian preserves the number of the particle along
$x$ direction.

\section{\label{Sec:4}Quantum dynamics of a quasispin dark state polariton}

\begin{figure}[tbp]
\includegraphics[bb=125 192 496 676, width=5 cm]{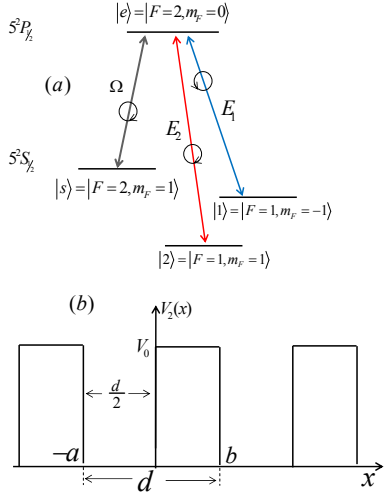}
\caption{(Color online) A possible experimental realization of the
tripod scheme in matter-light system in rubidium 87 (a) and an
illustration of the periodical potential for the second component of
DSP induced by the periodical magnetic field (b).} \label{fig2:2}
\end{figure}

To show the possibility that the dynamics of DSP mirrors the Bloch
oscillation in this atom--photon system, we consider the atomic medium with
light tuned to the rubidium ($^{87}$Rb) D1-line $5^{^{2}}S_{_{1/2}}%
\leftrightarrow 5^{^{2}}P_{_{1/2}}$ as shown in
Fig.~\ref{fig2:2}(a). The ground states contain two hyperfine ground
levels with $F=1$ and $F=2$, where $\left\vert 1\right\rangle $ and
$\left\vert 2\right\rangle $ correspond to the magnetic sublevels
(with $m_{F}=1$ and $-1$) of the $F=1$ hyperfine ground state and
$\left\vert s\right\rangle $ represents the hyperfine ground state
$\left\vert F=2,m_{F}=1\right\rangle $. In this case, the effective
magnetic moment $\mu _{1}^{eff}$ vanishes due to $\mu _{s}=\mu
_{1}$, which leads to the free propagation of the component $\Psi
_{1}$. Therefore, a wave packet with a Gaussian profile in space
\begin{eqnarray}
\Psi _{1}\left( x,z,0\right) =\prod_{\alpha
=x,z}\frac{1}{\sqrt[4]{2\pi \sigma _{\alpha }^{2}}}e^{-\frac{\alpha
^{2}}{4\sigma _{\alpha }^{2}}} \label{BOP1}
\end{eqnarray}%
at the initial time, will keep its shape in all directions. Since equation (\ref%
{BOP1}) has a pronounced peak with width $\sigma _{x}$ situated at moment $p_{x}=0$,
the wave-packet will always localize around the position $x=0$ along x-axis, but
the center of the wave-packet in $z$-direction moves to $z=v_{g}t$.

Different from $\Psi _{1}$, the second component $\Psi _{2}$ feels
potentials due to its nonzero effective magnetic moment. Obviously,
in the absence of the linear magnetic field, the period potential
gives rise to the Bloch bands. To make this discussion more quantitative,
we assume that $V_{2}\left( x\right) $ is the Kroning-Penney potential
which is formed by a periodic sequence of rectangular wells with amplitude
$V_{0}=79.15kHz$. As illustrated in Fig.~\ref{fig2:2}(b), the lattice
constant of the periodic array $d=8\mu m$ is equal to $a+b$. Within the
unit cell, the barrier region $0<x<a$, the well region $-b<x<0$. By
taking the effective transverse mass $m=7.9\times 10^{5}sm^{-2}$ for
the wavelength of the probe field $\lambda =795nm$, we obtain the width
$\Delta =74.9kHz$ of the ground energy band, and the gap interval
$E_{g}=266kHz$ between two lower bands. For a linear magnetic field
with magnitude $B_{1}=8.5\times 10^{4}\mu Gmm^{-1}$, the magnetic moments
$\mu_{2}=-4.64\times 10^{-24}JT^{-1}$ and $\mu _{s}=4.64\times
10^{-24}JT^{-1}$ allow us to find that $F_{2}d\ll E_{g}$. So the
ground band is well separated and one can neglect the interband
mixing induced by the static force $F_{2}$. By assuming that
neighboring wells are directly coupled, the dynamics of the DSP component $%
\Psi _{2}$ is described by the Hamiltonian%
\begin{eqnarray}  \label{BOP2}
H_{1}=\frac{-\Delta }{4}\sum_{n}\left( \left\vert n\right\rangle
\left\langle n+1\right\vert +h.c.\right) +dF_{2}\sum_{n}n\left\vert
n\right\rangle \left\langle n\right\vert  \notag \\
\end{eqnarray}%
in terms of Wannier state $\left\vert n\right\rangle $, which is
localized around position $x=nd$. It is useful to introduce the eigenstates $%
\left\vert \kappa\right\rangle $ (Bloch states) of Hamiltonian $H_{1}$ with $%
F_{2}=0$,
\begin{eqnarray}
\left\vert \kappa \right\rangle =\sqrt{\frac{d}{2\pi }}\sum_{n}\left\vert
n\right\rangle e^{in\kappa d}  \label{BOP3}
\end{eqnarray}%
which is the Fourier transform of the Wannier states. The tight-binding
Hamiltonian in Eq.(\ref{BOP2}) is diagonalized as
\begin{eqnarray}
H_{1}\left( k\right) =-\frac{\Delta }{2}\cos \left( \kappa d\right) +iF_{2}%
\frac{d}{d\kappa }  \label{BOP4}
\end{eqnarray}%
in the quasi-momentum representation with the quasi-momentum $\kappa \in %
\left[ -\pi /d,\pi /d\right] $ in the first Brillouin zone. Diagonalizing $%
H_{1}$ gives the Wannier-Stark ladder with energies $E_{m}=mdF_{2}$,
where $m$ is an integer. The Wannier--Stark state $\left\vert \psi
_{m}\right\rangle $ to the eigenvalue $E_{m}$ is given by%
\begin{eqnarray}
\left\vert \psi _{m}\right\rangle =\sqrt{\frac{d}{2\pi }}\int d\kappa e^{-i%
\left[ \Delta \sin \left( \kappa d\right) /\left( 2dF_{2}\right) +md\kappa %
\right] }\left\vert \kappa \right\rangle  \label{BOP5}
\end{eqnarray}%
In the Wannier-state representation, we obtain
\begin{eqnarray}
\left\vert \psi _{m}\right\rangle =\sum_{n}J_{n-m}\left( \frac{\Delta }{%
2dF_{2}}\right) \left\vert n\right\rangle  \label{BOP6}
\end{eqnarray}%
where $J_{n-m}$ is the Bessel functions of the first kind and of integer
order $n-m$.

To show the wave-particle duality of the second DSP component, we
now discuss the quantum dynamics under the assumption that an
initially Gaussian wave packet
\begin{eqnarray}
\left\vert \varphi\left( 0\right) \right\rangle =\sum_{n}\left( \frac{%
d^{2}}{2\pi \sigma ^{2}}\right) ^{1/4}\exp \left[ -\frac{n^{2}d^{2}}{4\sigma
^{2}}\right] \left\vert n\right\rangle  \label{BOP7}
\end{eqnarray}%
is launched into the lattice with a center value at $x_{0}=0$ and width $%
\sigma $. Since the time-evolution operator is diagonal in Wannier--Stark
states, each element $U_{\mu \nu }\equiv \left\langle \mu \right\vert
U\left( t\right) \left\vert \nu \right\rangle $ of the time-evolution
operator $U\left( t\right)$ in the Wannier-state representation has the form
\begin{eqnarray}
U_{\mu \nu }=J_{\nu -\mu }\left( \frac{\Delta \sin \alpha }{\omega _{_{B}}}%
\right) e^{-i\mu \omega _{_{B}}t}e^{-i\left( \nu -\mu \right) \left(
\omega _{_{B}}t-\pi \right) /2},  \label{BOP8}
\end{eqnarray}%
where $\alpha =\omega _{_{B}}t/2$ and $\omega _{_{B}}=dF_{_{2}}$.
The argument of $J_{\nu -\mu }$ oscillates in time. The
state at arbitrary time can be expressed as $\left\vert \varphi\left(
t\right) \right\rangle =\sum_{n}f_{n}\left( t\right) \left\vert
n\right\rangle $. As the Gaussian wave packet is broad initially, i.e.
$\sigma /d$ $\gg 1$, the coefficient $f_{n}\left( t\right) $ is
approximately given by
\begin{eqnarray}
f_{n}\left( t\right) =\frac{e^{-d^{2}\left( n-n_{t}\right) ^{2}/4\sigma ^{2}}%
}{\sqrt[4]{2\pi \sigma ^{2}/d^{2}}}e^{-in\omega _{_{B}}t+i\Delta \sin \left(
\omega _{_{B}}t\right) /(2\omega _{_{B}})}  \label{BOP9}
\end{eqnarray}%
where $n_{t}=x_{c}\left( t\right) /d$. Here, the motion
of the wave packet center along $x$-direction%
\begin{eqnarray}
x_{c}\left( t\right) =\frac{\Delta }{2F_{2}}\left[ \cos \left(
\omega _{_{B}}t\right)- 1\right] \label{BOP10}
\end{eqnarray}%
reduces the quantum mechanical dynamics to the classical
trajectories. Equation (\ref{BOP10}) implies that the DSP component
$\Psi _{2}$ experiences a harmonic motion around the initial
position $x_{0}$ with angular frequency $\omega _{_{B}}$ and
amplitude $A=\Delta /\left( 2F_{2}\right)$. As expected from the
semiclassical picture, the band width $\Delta $ equals the product
of the total spatial extension $2A$ and the static force $F_{2}$.
Besides the trajectory of the wave packet, equation~(\ref{BOP9})
also gives a description on the variation of the particle's
wavenumber $\kappa =\omega _{_{B}}t/d=F_{_{2}}t$ in the
semiclassical picture. Since $z=v_{g}t$ is the relation between the
center of the wave packet along the z direction and
the time, we achieve the spatial Bloch oscillation%
\begin{eqnarray*}
x\left( z\right) =A\left[ 1-\cos \left( \zeta z\right) \right]
\end{eqnarray*}
with spatial period $2\pi /\zeta $. Here, $\zeta =dF_{2}/v_{g}$.
\begin{figure}[tbp]
\includegraphics[bb=4 14 316 281, width=7cm]{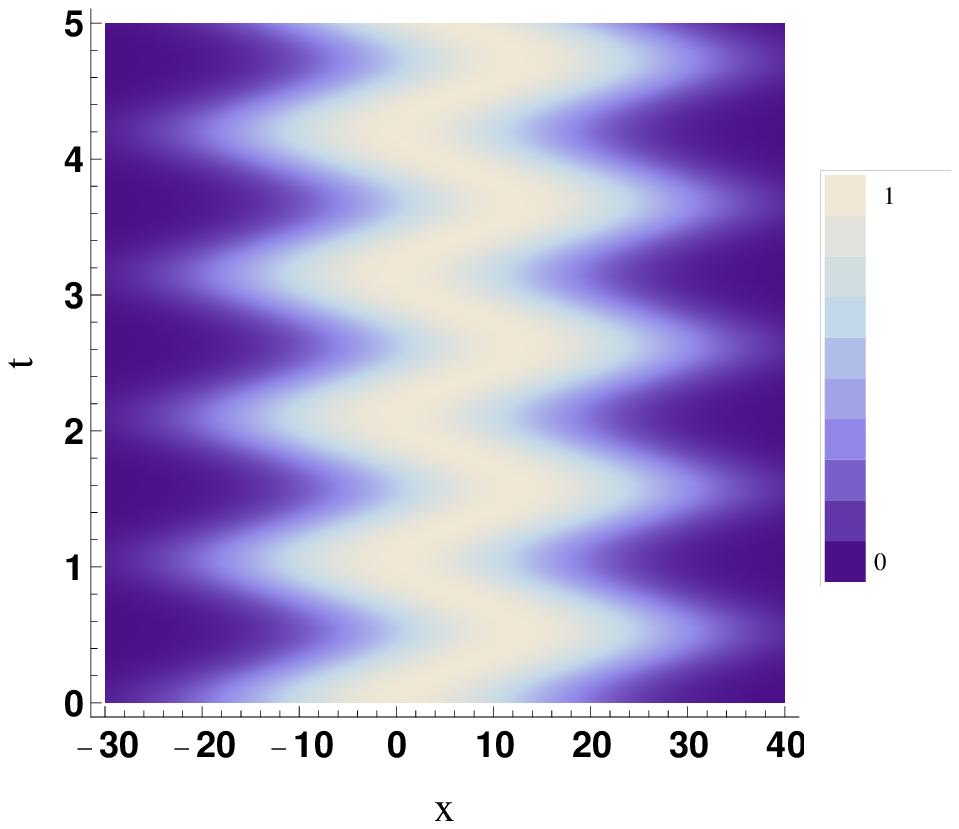}
\includegraphics[bb=6 18 293 272, width=7cm]{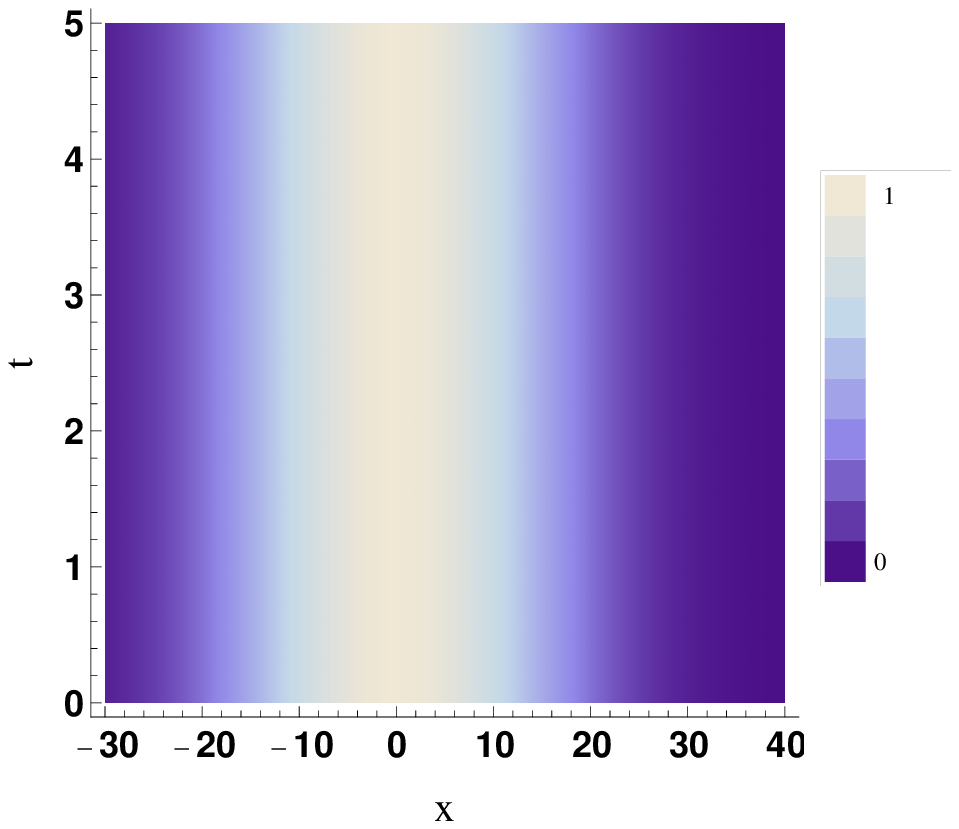}
\caption{(Color online) Density plots of the norm square of the wave
function for a DSP with two components arising from an extended
Gaussian distribution. The probabilities for the components $\Psi_{2}$
(a) and $\Psi_{1}$ (b) is plot as a function of x in units of lattice
constant $d$ and t in units of millisecond, where the initial states
for both components are taken to be the same with width $0.1$mm.}
\label{fig3:DSP}
\end{figure}
In Fig.~\ref{fig3:DSP}, we plot the probabilities for the components $%
\Psi_{1}$ and $\Psi_{2}$ versus $x/d$ and $t$, where the initial states
of both components are identical broad Gaussian wave packets. One can see
that two components behave differently in the transverse direction though
the wave packets for both components keeps it shape along z direction
with an unchanged variance. The trajectory of component $\Psi_{1}$ is
always localized at its initial position, which means this component
appears like a free particle. However, component $\Psi_{2}$ presents
a back and forth behavior around an equilibrium point with a constant
amplitude and a constant frequency, so it acts as a particle moving
through a quadratic potential.

\section{\label{Sec:5}Conclusion}

We study an optical-controlled cloud of identical atoms in a
washboard magnetic field applied along $z$-direction. The probe and
control laser beams drive three hyperfine ground states to a common
excited state, which form a tripod configuration. The quasispin DSP
is excited in the EIT condition. The equation for the space-time
evolution of this quasispin DSP shows that each component is subject
to its corresponding effective potential induced by the steady
atomic response in the external spatial-dependent field. By taking
the value of parameters from the experiment data in rubidium
D1-line, it is found that one component $\Psi _{1}$ of the DSP acts
as a free particle. The behavior of the other component $\Psi _{2}$
is analyzed via the single band and tight-binding approximation. By
considering the time evolution of the broad Gaussian state for the
component $\Psi _{2}$, the particle nature of component $\Psi _{2}$
is shown by the trajectory of the wavepacket, which undergoes a
periodic motion with angular frequency proportional to the steady
force. Besides, its momentum grows linearly in time. This periodic
motion is termed as Bloch oscillations. The oscillation amplitude
and period are controlled by the intensity of the control field
and the magnitude of the magnetic fields.

It should be pointed out that there are many differences between our
present scheme and that in Ref.~\cite{ZhangA10}: 1) The atomic
configurations are different. Lambda type is studied in
Ref.~\cite{ZhangA10}, however tripod pattern is studied in our
manuscript. From the technical point of view, tripod atoms proved to
be robust systems for ``engineering'' arbitrary coherent
superpositions of atomic states~\cite{VewPRL91} using an extension
of the well known technique of stimulated Raman adiabatic passage.
2) The double-EIT effect has been used in our manuscript. Double EIT
is an important phenomenon for quantum information processing and
quantum computing. 3) In our paper, different types of dark
polaritons are demonstrated to possess different effective magnetic
moments. 4) The predicted phenomena in our manuscript are more
general. It could be found that only one dark polariton is excited
in Ref.~\cite{ZhangA10}, so only one trajectory is found inside the
atomic medium. However, in our paper, two dark polaritons are
excited by one probe beam and a spatial bifurcation and dynamics of
these dark polaritons is obtained inside the atomic medium. 5) Our
approach naturally shows the multiple degrees of freedom of photons
and the role of quantum coherence. As one can find that when the
single photon with a superposition of two orthogonal components of
polarization is incident on the atomic medium, one obtain a
superposition state of two DSPs relating to different spin-states of
this incident single photon. Besides, our investigation provides a
way to control the dynamics of DSPs. In addition to the external
fields, atomic energy levels can be used as a way to adjust the
motion of DSP, for example, one can choose the atomic energy level
to make the DSP have vanished effective magnetic moment. The
presence of the first component unaffected by the washboard
potential might be used to eliminate the distortions or aberrations
in future.

This work is supported by the Program for New Century Excellent Talents in
University (NCET-08-0682), NSFC No.~11105050, and No. 11074071, NFRPC
2012CB922103, PCSIRT No.~IRT0964,
the Key Project of Chinese Ministry of Education (No.~210150), the Research
Fund for the Doctoral Program of Higher Education No. 20104306120003, Hunan
Provincial Natural Science Foundation of China(11JJ7001), and Scientific
Research Fund of Hunan Provincial Education Department (No.~11B076).We
thanks H. R. Zhang for useful discussions.

\end{document}